# Deep Learning-based Online Alternative Product Recommendations at Scale

**Mingming Guo, Nian Yan, Xiquan Cui, San He Wu, Unaiza Ahsan, Rebecca West and Khalifeh Al Jadda**

The Home Depot, Atlanta, GA, USA

{mingming_guo, nian_yan, xiquan_cui, san_h_wu, unaiza_ahsan, rebecca_west, khalifeh_al_jadda}@homedepot.com

## Abstract

Alternative recommender systems are critical for ecommerce companies. They guide customers to explore a massive product catalog and assist customers to find the right products among an overwhelming number of options. However, it is a non-trivial task to recommend alternative products that fit customers' needs. In this paper, we use both textual product information (e.g. product titles and descriptions) and customer behavior data to recommend alternative products. Our results show that the coverage of alternative products is significantly improved in offline evaluations as well as recall and precision. The final A/B test shows that our algorithm increases the conversion rate by 12% in a statistically significant way. In order to better capture the semantic meaning of product information, we build a Siamese Network with Bidirectional LSTM to learn product embeddings. In order to learn a similarity space that better matches the preference of real customers, we use co-compared data from historical customer behavior as labels to train the network. In addition, we use NMSLIB to accelerate the computationally expensive kNN computation for millions of products so that the alternative recommendation is able to scale across the entire catalog of a major ecommerce site.

## 1 Introduction

Recommender systems are pervasive in ecommerce and other web systems (Zhang et al. 2019). Alternative product recommendation is an important way to help customers easily find the right products and speed up their buying decision process. For example, if a customer is viewing a "25.5 cu. ft. Counter Depth French Door Refrigerator in Stainless Steel", she may also be interested in other french door refrigerators in different brands but with similar features such as capacity, counter depth, material, etc.

There are two main ways to obtain an alternative product list for a given product. First is a content-based recommendation approach. If two products have similar attributes or content so that one can be replaced by the other, we can consider them as alternative products. Word2vec has been used to learn item embeddings for comparing item similarities (Caselles-Dupre, Lesaint, and Royo-Letelier 2018). However, this unsupervised learning process does not guarantee the embedding distance is consistent with customers' shopping preference. The second way is to leverage customer behavior to find alternative products in the style of item-to-item collaborative filtering (Linden, Smith, and York 2003). If customers frequently consider two products together, one product can be recommended as an alternative for the other. Unfortunately, this approach has a cold start problem.

In this work, we formulate the recommendation problem into a supervised product embedding learning process. To be specific, we develop a deep learning based embedding approach using Siamese Network, which leverages both product content (including title and description) and customer behavior to generate Top-N recommendations for an anchor product. Our contributions are as follows:

- Recommend alternative products using both product textual information and customer behavior data. This allows us to better handle both the cold start and relevancy problems.

- Use a Bidirectional LSTM structure to better capture the semantic meaning of product textual information.

- Build a Siamese Network to incorporate co-compared customer behavior data to guide the supervised learning process and generate a product embedding space that better matches customer's preference.

- Our model outperforms baselines in both offline validations and an online A/B test.

## 2 Problem Formulation

We have the textual information $T = \{x_1, ..., x_N\}$ (a concatenation of product title and description) of a catalog of products $P = \{p_1, ..., p_N\}$ to make recommendations. The goal of the alternative recommendation is to learn an embedding projection function $f_w$ so that the embedding of an anchor product that is viewed by a customer $f_w(x_a)$ is close to the embeddings of its alternatives $f_w(x_r)$. In this paper, we use the cosine similarity between the embeddings of $f_w(x_a)$ and $f_w(x_r)$ as the energy function.

$$E_w = \frac{\langle f_w(x_a), f_w(x_r) \rangle}{\|f_w(x_a)\| \|f_w(x_r)\|} \quad (1)$$

The problem is how to learn a function as the embedding projection function $f_w$ to better capture the semantic meanings of the product textual information and project a sequence of tokens $x_i$ into an embedding vector of size d. The total loss over the training data set $X = \{x_a^{(i)}, x_r^{(i)}, y^{(i)}\}$ is given by

$$L_w(X) = \sum_{i=1}^{M} L_w^{(i)}(x_a^{(i)}, x_r^{(i)}, y^{(i)}) \quad (2)$$

where the instance loss function $L_w^{(i)}$ is a contrastive loss function. It consists of a term $L_+$ for the positive cases ($y^{(i)} = 1$), where the product pair are alternative to each other. In addition, it consists of a term $L_-$ for the negative cases ($y^{(i)} = 0$), where the product pair are not often considered together by customers.

$$L_w^{(i)} = y^{(i)} L_+(x_a^{(i)}, x_r^{(i)}) + (1 - y^{(i)}) L_-(x_a^{(i)}, x_r^{(i)}) \quad (3)$$

The loss functions for the positive and negative cases are given by:

$$L_+\left(x_a^{(i)}, x_r^{(i)}\right) = |1 - E_w| \quad (4)$$

$$L_-\left(x_a^{(i)}, x_r^{(i)}\right) = \begin{cases} |E_w| & if\ E_w > 0 \\ 0 & otherwise \end{cases} \quad (5)$$

Based on the loss function, the problem is how to build a network that can learn part of the product information that is important for customers and project a product to the right embedding space that is consistent with customers' preference.

## 3 Deep Learning Embedding Approach

**Textual Data and Co-compared Data**

Product Information: From the ecommerce site catalog data, we extract the product ID, product title and description as the raw textual data with an example in Table 1.

| Product ID | Product Title | Product Description |
|---|---|---|
| '12345678' | 60 Gal. Electric Air Compressor | This compressor offers a solid cast iron, twin cylinder compressor pump for extreme durability. It also offers 135 psi maximum pressure and air delivery 11.5/10.2 SCFM at 40/90 psi. |

Table 1. Product Textual Data

Co-compared Data: Customers can select several products on a search result page for co-comparison to verify how they are similar and different based on their features. Those products are considered alternative to each other. The co-compared is a strong signal of the similarity between products within same product taxonomy. We extract co-compared data from clickstream to create the training data. Some examples of the co-compared data are shown in Table 2.

| Product ID_1 | Product ID_2 | Co-compared |
|---|---|---|
| '12345678' | '87654321' | 1 |
| '32187654' | '54321876' | 1 |

Table 2. Co-compared Example

**Siamese Network with Bidirectional LSTM**

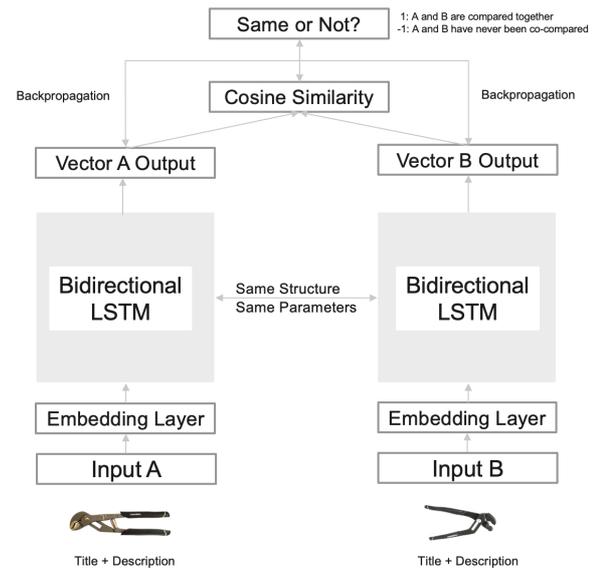

Fig. 1 Siamese Network with Bidirectional LSTM

We build a Siamese Network (Bromley et al. 1994) with Bidirectional LSTM (Graves and Schmidhuber 2005) components to learn and generate embeddings for all products. The product embedding space better captures the semantic meaning of the product textural information and customer preferences. Textual data are in a sequential format and the order of the texts matters for the network. We choose Bidirectional LSTM to learn representation in both directions from the input sequences. We use Keras (Chollet et al. 2015) with TensorFlow to build and train the network. We choose RMSprop (Hinton et al. 2012) as the optimizer. The loss function is the binary cross entropy. The network architecture is shown in Figure 1.

**Creating Training Data by Sampling**

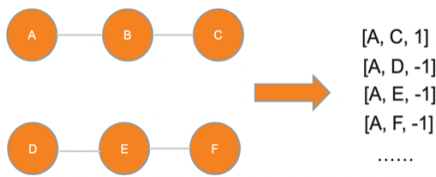

Fig. 2 Connected Graphs

Positive and negative sampling: we form a connected graph from the co-compared product pairs. For example, if product A and B are co-compared and product B and C are co-compared, then (A, B, C) forms a connected graph. If products D and E are co-compared and E and F are co-compared, then (D, E, F) forms a connection graph. We create positive samples for each product by randomly sampling another product within the same set, e.g. [A, C, 1]. We also create negative samples for each product in a connected graph by randomly sampling a different connected graph first, then randomly sampling a product in that graph, e.g. [A, D, -1], as shown in Figure 2. We use '-1' for illustration of the negative label and replace it with '0' in the final training data.

| Number of training data records | Number of products involved |
|---|---|
| 331900 | 65684 |

Table 3. Training Data Statistics

The negative sampling space is much larger than the positive sampling space because only a small number of products are frequently co-compared together by our customers. Thus, for each anchor product, we sample more negative samples than positive samples. Based on our experiments and empirically analysis, for each positive sample, three negative samples are created which gives the best performance on the validation loss when training the model. The statistics of the training data is shown in Table 3.

**Training the Model and Generating Embeddings**

The Siamese Network training process takes about 10 hours to converge. The next step is to load the best model weights to generate product embeddings. Specifically, from the Siamese Network, we remove the last cosine similarity layer and the second input branch which processes the second product of the product pairs. We only use the Embedding layer and the Bidirectional LSTM layer. The final result is the concatenation of the hidden state of the product title and the hidden state of the product description.

**Scalable Recommendation Generation**

We generate millions of embeddings based on product titles and descriptions. For each product, the task is to compute distances with the rest of millions of product embeddings using a similarity metric, e.g. cosine similarity, and rank the similarity scores from higher to lower to get the Top-N recommended products. According to the detailed analysis from (Aumüller et al. 2019), we choose NMSLIB (Boytsov et al. 2016) library to conduct heavy kNN computations because it has high performance in both recall and queries per second.

## 4   Performance Evaluation

In this section, we describe how we evaluate the effectiveness and efficiency of our deep learning model with offline evaluation and online A/B test. We use our production data to validate the results since this is a unique case for us. We did not find exact similar open data set with similar customer behaviors that can be used for our evaluation.

**Algorithms:**

**1) Baseline 1: Attributed Based**

This baseline algorithm uses product attributes to generate recommendations. The attributes contain numerical and categorical data. The categorical features are converted into numerical format using one-hot encoding. The distance between two products is computed using cosine similarity. This is the content-based method we compare with.

**2) Baseline 2: Frequently Compared**

This baseline algorithm uses the actual customer co-compared data. The recommendations are ranked by the co-comparison counts. Due to the cold-start problem, many products in the catalog do not have such recommendations even we create labels from the co-compared data. This is the collaborative filtering method we compare with since it's based on item-to-item relationships built by customer browsing behaviors.

### 3) Proposed: Deep Learning Based

For Deep Learning Based, we choose 0.8 as the cutting threshold for the cosine similarity score. This threshold is selected and validated based on the judgement from our human expert validators after they examine thousands of random sampled anchors from the catalog data and the recommendations generated from our model. We only keep the recommendations that have at least 0.8 similarity with each anchor product.

## Offline Evaluation:

### 1) Precision and Recall:

|  | Precision | | | Recall | | |
|---|---|---|---|---|---|---|
|  | Top 1 | Top 5 | Top 10 | Top 1 | Top 5 | Top 10 |
| **Attribute Based** | 0.23% | 0.13% | 0.10% | 0.15% | 0.34% | 0.33% |
| **Frequently Compared** | 0.75% | 0.51% | 0.47% | 0.51% | 0.93% | 1.02% |
| **Deep Learning Based** | **1.47%** | **0.81%** | **0.61%** | **0.91%** | **2.08%** | **2.59%** |

Table 4. Precision and Recall with Raw Sessions

**Comparison 1:**

Two weeks of actual customer purchase data from clickstream data is used to evaluate the performance of all 3 algorithms based on precision and recall. We only use the sessions with purchase behavior for evaluation. In this comparison, we use the raw data regardless if each session has all two baselines. This is a fair comparison since not all anchors can be covered by both algorithms. For example, a product may not have the same set of attributes as other products so this product cannot be covered by Attributed Based algorithm. This is because there are vast variants of similar products without same set of attributes. Another scenario is that this product has never been compared with other products by our customers so this product cannot be covered by Frequently Compared algorithm. For the Deep Learning Based, we compare its recommendations with the purchased items. Table 4 shows our algorithm performs much better than the baseline algorithms for all top 1, 5, and 10 items precision and recall scores, especially for precision top 1, recall top 5 and top 10. The main reasons are: i) Frequently Compared recommends co-compared products by customers and only covers small sets of products; ii) Attributed Based approach has a higher coverage but a lower relevancy.

**Comparison 2:**

|  | Precision | | | Recall | | |
|---|---|---|---|---|---|---|
|  | Top 1 | Top 5 | Top 10 | Top 1 | Top 5 | Top 10 |
| **Attribute Based** | 0.21% | 0.13% | 0.12% | 0.16% | 0.31% | 0.34% |
| **Frequently Compared** | **2.48%** | **1.81%** | **1.75%** | **1.65%** | **2.65%** | 2.76% |
| **Deep Learning Based** | 1.72% | 0.90% | 0.67% | 1.09% | 2.33% | **2.85%** |

Table 5. Precision and Recall with Filtered Sessions

In this comparison, we select sessions that have both Attributed Based and Frequently Compared. Table 5 shows our Deep Learning Based still performs much better than Attributed Based but not Frequently Compared. The reason is that the label we used to train our model is from co-compared data, so our model has the upbound from Frequently Compared's performance. This experiment validated our hypothesis.

**2) Coverage**: The anchor coverages of all the algorithms are also computed. Compared with Attributes Based approach, our Deep Learning Based approach increases 49.7% of the anchor coverage. Compared with Frequently Compared approach, our Deep Learning Based approach also increases the anchor coverage by 36.3%. Since most of our products have titles and descriptions, so our Deep Learning Based significantly boosts the coverage of anchor products from our catalog to have good recommendations.

## Online A/B Testing:

**Conversion Rate**: The A/B test was run for three weeks and success was measured using conversion rate. Conversion rate is the number of purchases divided by number of visits which captures the similarity between anchor and recommendations. Our deep learning model outperforms the existing hybrid algorithm which combined Attribute Based and Frequently Compared with a **12%** higher conversion rate. This is a very successful test for our business. We're implementing the deep learning algorithm on our production site.

## 5 Related Work

The traditional method for recommender systems is content-based recommendations (Lops et al. 2011). This method can handle the cold start problem well. Collaborative Filtering is another method based on user behaviors. For example, Matrix Factorization (Koren et al. 2009) is a widely used method for collaborative filtering. Our two baseline algorithms, one is considered as content-based and the other is considered as collaborative filtering using user behavior data with the co-compared format. Deep learning now has been widely used not only in the academic community, but also in industrial recommender system settings, such as Airbnb's listing recommendations (Grbovic and Cheng 2018) and Pinterest's recommendation engine (Ying et al. 2018). Most of recent deep learning papers (e.g., Wang et al. 2019; Ebesu, Shen, and Fang 2018) have been focused on personalized recommendations. A deep network using Siamese architecture with character-level Bidirectional LSTM is proposed for job title normalization (Neculoiu et al. 2016). Mueller and Thyagarajan (2016) also presented a Siamese adaptation of the LSTM to learn sentence embedding. However, this work needs human annotated labels while our labels are extracted from clickstream data. Our work more focuses on providing alternative recommendations by learning product embedding from product textual data and customer signals.

## 6 Conclusion

Recommender Systems are core functions for online retailers to increase their revenue. To help customers easily find alternative products in an automated way, we develop a deep learning approach to generate product embeddings based on a Siamese Network with Bidirectional LSTM. We extract co-compared data from customer clickstream and product textual data to train the network and generate the embedding space. Our approach significantly improves the coverage of similar products as well as improving recall and precision. Our algorithm also shows promising results on conversion rate in an online A/B test.